\documentstyle[aps,epsfig]{revtex}

\begin{document}
\draft

\title{\null\vspace{0.2cm}\null\hspace{10cm}TUIMP-TH-97/93\\
\null\vspace{0.5cm}
\Large\bf{Ward-Takahashi identity and dynamical mass generation\\
 in Abelian gauge theories}}
%\null\vspace{0.8cm}
\author{ Kun  Shen$^{a,b,c}$~~ and~~ Yu-Ping Kuang$^{a,b}$}
\address{ a. CCAST(World Laboratory), P.O. Box 8730, Beijing 100080, China\\
b. Institute of Modern Physics, Tsinghua University, Beijing, 100084, China \\
c. Wuhan Research Institute of Posts and Telecommunications,
Wuhan, 430074, China\thanks{ Mailing address;  ~~
E-mail addresses:~~shenk@yky-20.wript.edu.cn,~~~ypkuang@tsinghua.edu.cn} }

\maketitle

\null\vspace{0.5cm}
\begin{abstract}
We derive Ward-Takahashi identities including composite fields in
Abelian gauge theories and the matching condition between the
elementary field description and the composite field description. 
With these we develop an approach to dynamical 
symmetry breaking in Abelian gauge theories including the study of the 
dynamically generated masses of the gauge boson, the fermions and the 
composite Higgs field.
The Cornwall-Norton, Jackiw-Johnson and Schwinger models are taken as examples
of the application. 
The obtained gauge boson masses are in agreement with the existing results. 
In this approach, we are able to further obtain new results for the
mass of the composite Higgs boson and the Goldstone boson decay constant 
$F_{\pi}$.
\end{abstract}

%\newpage

%\twocolumn

%\narrowtext

\section{Introduction}
\label{sec:intro}

Dynamical symmetry breaking plays an important role in particle physics.
In strong interaction,  chiral symmetry is realized in this pattern 
and many low-energy phenomena are determined by  the dynamically broken
chiral symmetry. On the other hand, dynamical symmetry breaking 
provides an alternative mechanism of mass generation in the electroweak 
theory. In this approach, all interactions are governed by the gauge principle
so that the number of parameters can be reduced, and the problems of
triviality and unnaturalness due to elementary Higgs fields do not occur.
Because of these attractive properties, many dynamical electroweak
symmetry breaking models have been proposed \cite{1}. 

The reason that a gauge field can obtain a mass dynamically was first given by 
Schwinger\cite{s}. He made a crucial observation that if the 
polarization tensor of vector mesons has a pole at $q^2=0$, the 
gauge symmetry will be spontaneously broken and the 
vector mesons acquire a mass. This is called Schwinger mechanism or the 
dynamical Higgs mechanism\cite{H}.
He also gave an example in two-dimensional quantum electrodynamics.
In the Schwinger model, the symmetry breaking leading to massive vector field 
is due to anomalies, rather than bound state Goldstone bosons.
Inspired by the pioneering work of Numbu concerning the gauge 
invariance of the BCS theory of super-conductivity\cite{Nambu}, 
Jackiw and Johnson and Cornwall and Norton extended the work of 
Numbu and Jona-Lasinio to dynamical gauge 
symmetry breaking\cite{CN,JJ}. They constructed 
two simple Abelian gauge theoretical models and showed that there are 
nontrivial self-consistent solutions with the vector mesons acquiring masses
dynamically.

Since dynamical symmetry breaking is a nonperturbative phenomenon,
it is  difficult to study it quantitatively and most of the investigations 
on dynamical symmetry breaking are based on the analogy of the low-energy 
phenomenology and current algebra. In order to study dynamical symmetry
breaking from the first principles of the fundamental theory, it is 
interesting to explore new approaches.

Recently, we derived Ward-Takahashi identities including composite fields  
and applied them to the study of chiral symmetry breaking and the properties 
of fermion bound states\cite{sq1}.  In some  model investigations, this method 
can be applied to obtain the mass spectra of both fermion and bound states 
bosons, and the results are in agreement with those from other methods. It is 
also convenient to used it to study the phase structure, mass spectrum and 
PCAC in the presence of explicit breaking terms\cite{sq2}. In this paper, we 
generalize this method to the case of dynamical breaking of gauge symmetry 
and study the mechanism of gauge boson mass generations and the
properties of the composite bosons. 

As a test of this approach, we take the simple and well understood 
Cornwall-Norton, Jackiw-Johnson and Schwinger models as the
application. We shall calculate the dynamically generated masses of the
gauge bosons in these models to test our method and calculate new results
related to the composite Higgs boson to show the ability of this new approach.

In Sec. \ref{sec:CN}, we study Cornwall-Norton model. As there are bound 
states in dynamical symmetry breaking, we introduce composite external sources 
in the generating functional. From  gauge symmetry, we derive the  
Ward-Takahashi identity with composite fields in this model, and with which 
we study the dynamical symmetry breaking and the mass spectra of the fermions, 
the vector mesons and bound state boson.
It is shown that when the gauge symmetry is dynamically broken, one of the 
vector mesons acquires a mass. This is in agreement with the picture of
the Schwinger mechanism. In the platform approximation, vector meson 
mass is identical to those given by Cornwall and Norton.
For a dynamically broken gauge theory, we can either describe it by
extracting the composite field degree of freedom in the effective action or
describe it only by elementary fields in which the bound states reside
implicitly in the effective action. These two descriptions should be 
equivalent. From this requirement we derive a matching condition between 
two descriptions.
Some physical quantities like $F_{\pi}$ can be expressed in terms of basic 
quantities by means of the matching condition.

In Sec. \ref{sec:JJ}, we apply the approach to the 
Jackiw-Johnson model. Considering that in two space-time dimensions anomalies 
can provide a pole in the vacuum polarization tensor of the vector boson, we 
apply this approach to the Schwinger model in Sec. \ref{sec:S}, and find that 
the vector boson does acquire a mass from the anomaly, which is identical to 
other approaches. Some comments and conclusions are presented in 
Sec. \ref{sec:D}. 

\section{ The Cornwall-Norton Model}
\label{sec:CN} 
\subsection{ Ward-Takahashi Identity}

The Cornwall-Norton model contains two fermions with equal bare masses and two 
massless Abelian gauge fields. The gauge fields couple to the fermions
via different currents. There are two $U(1)$ ($O(2)$) symmetries in this model. 
The first one is related to the rotation within the two fermion fields. This 
$O(2)$ symmetry is dynamically broken which makes the associated gauge field 
massive. The other one is an unbroken $U_V(1)$ symmetry and the associated 
gauge field remains massless. The Cornwall-Norton model is described 
by the following Lagrangian density
\begin{eqnarray}
{\cal L}=&&\bar\psi(i\gamma\cdot\partial- m_0)\psi
 -{1\over 4}F_{\mu\nu}F_{\mu\nu}
-{1\over 4}G_{\mu\nu}G_{\mu\nu}+g\bar\psi\gamma_{\mu}\psi A_{\mu}+g'\bar\psi\gamma_{\mu}\tau_2\psi
B_{\mu},
\end{eqnarray}
where 
\begin{mathletters}
\begin{eqnarray}
 &&F_{\mu\nu}=\partial_{\mu}A_{\nu}-\partial_{\nu}A_{\mu},\\
&&G_{\mu\nu}=\partial_{\mu}B_{\nu}-\partial_{\nu}B_{\mu},
\end{eqnarray}
\end{mathletters}
 $\psi $ represents the two fermion fields $\psi_1$ and $\psi_2$, 
$\tau_2$ is the Pauli matrix and $A_{\mu}, B_{\mu}$ are Abelian gauge fields.

The Lagrangian density is locally invariant under the following 
transformations
\begin{mathletters}
\begin{eqnarray}
&&\delta \psi(x)=i(\alpha(x)+\tau_2\beta(x))\psi(x),\\
&&\delta A_{\mu}(x)={1\over g}\partial_{\mu}\alpha(x),\\
&&\delta B_{\mu}(x)={1\over g'}\partial_{\mu}\beta(x),
\end{eqnarray}
\end{mathletters}
\noindent 
where $\alpha(x), \beta(x)$ are the infinitesimal parameters
of $U_V(1)$ and $O(2)$ respectively.
The $U_V(1)$ symmetry is related to the fermion number conservation, 
and it remains unbroken.

Considering that the $O(2)$ symmetry is dynamically broken, there can be
fermion pair condensates and bound states in the broken phase.  In order to 
describe the fermion pair condensates,  we introduce the external source
$K_a$ coupling to the composite operator $\bar{\psi}\tau_a\psi$ in the 
generating functional\cite{sq1} 
\begin{eqnarray}
Z[J]\equiv &&e^{iW[J]}\nonumber\\
    =&&\int D\bar\psi D\psi DA_{\mu} DB_{\mu} 
\exp \biggl(i\int d^4x[{\cal L}+J_{\mu}B_{\mu}+I_{\mu}A_{\mu}+\bar\eta\psi
+\bar\psi\eta+\bar\psi\tau_a\psi K_a]\biggr),
\end{eqnarray}
where $J$ denotes the abbreviation of all external sources
$(J_{\mu},I_{\mu}, \bar\eta,\eta, K_a)$.
 
Define 
\begin{mathletters}
\begin{eqnarray}
&&{\delta W[J]\over \delta J_{\mu}(x)}=B^{\mu}_c(x),\\
&&{\delta W[J]\over \delta I_{\mu}(x)}=A^{\mu}_c(x),\\
&&{\delta W[J]\over \delta \bar\eta(x)}=\psi_c(x),\\
&&{\delta W[J]\over \delta \eta(x)}=-\bar\psi_c(x),\\ 
&&-{1\over i}{\delta \over \delta \eta(x)}\tau_a
{\delta \over \delta \bar\eta(x)}W[J]=G_a(x).
\end{eqnarray}
\end{mathletters}
From the above equation, one finds
\begin{eqnarray}
{\delta W[J]\over \delta K_a(x)}&=&G_a(x)+\bar\psi_c(x)\tau_a\psi_c(x).
\end{eqnarray}

Making the Legendre transformation, we get
\begin{eqnarray}
\Gamma [\phi]&=&W[J]-\int d^4x [\bar\psi_c(x)\eta(x)+\bar\eta(x)\psi_c(x)+J_{\mu}(x)B^{\mu}_c(x)+
\nonumber\\
&&+I_{\mu}(x)A^{\mu}_c(x)+ +(G_a(x)+\bar\psi_c(x)\tau_a\psi_c(x))K_a(x)].
\label{eq14}
\end{eqnarray}
Then we have 
\begin{mathletters}
\begin{eqnarray}
{\delta \Gamma [\phi]\over \delta \psi_c(x)}&=&\bar\eta(x)
+\bar\psi_c(x)\tau_a K_a(x),\\
{\delta \Gamma [\phi]\over \delta \bar\psi_c(x)}&=&-\eta(x)
-\tau_a\psi_c(x) K_a(x),\\
{\delta \Gamma [\phi]\over \delta B^{\mu}_c(x)}&=&-J_{\mu}(x),\\
{\delta \Gamma [\phi]\over \delta A^{\mu}_c(x)}&=&-I_{\mu}(x),\\
{\delta \Gamma [\phi]\over \delta G_a(x)}&=&-K_a(x).
 \end{eqnarray}
\end{mathletters}

Under the $O(2)$ transformation, we can obtain the following generating
equation of the Ward-Takahashi identities
\begin{eqnarray}
&&{\delta \Gamma [\phi]\over \delta \psi_c(x)}i\tau_2\psi_c(x)
+\bar\psi_c(x)i\tau_2{\delta \Gamma [\phi]\over \delta
\bar\psi_c(x)}+{1\over g'}\partial^{\mu}{\delta \Gamma [\phi]\over \delta B^{\mu}_c(x)}
-2\epsilon_{2ab}\sigma^a_c(x){\delta \Gamma [\phi]\over \delta \sigma^b_c(x)}
=0,
\label{c1}
\end{eqnarray}
where we have introduced a local field $\sigma^a(x)\equiv aG^a(x) $
to describe the bound state degree of freedom $G^a(x)$ with $a$ a 
dimensionful coefficient making $dim.\sigma^a=1$. From (\ref{c1})
we can get the Ward-Takahashi identity for the two-point
one-particle-irreducible (1PI) vertex.
As the two-point 1PI vertex is related to the mass spectrum, we can get the 
mass spectra of fermion and gauge field.

%\widetext
\subsection{ Mass Spectrum}
Taking derivatives of (\ref{c1}) with respect to $\psi_c(y)$ and 
$\bar\psi_c(z)$, one gets
\begin{eqnarray}
&&\bar\psi_c(x)i\tau_2{\delta^3 \Gamma [\phi]\over \delta \bar\psi_c(z)
\delta \psi_c(y)\delta \bar\psi_c(x)}
-\delta (x-z)i\tau_2{\delta^2 \Gamma [\phi]\over \delta \psi_c(y)
\delta \bar\psi_c(x)}-\delta (x-y){\delta^2 \Gamma [\phi]\over \delta\bar \psi_c(z)
\delta \psi_c(x)}i\tau_2+
\nonumber\\
&&+{\delta^3 \Gamma [\phi]\over \delta \bar\psi_c(z)
\delta \psi_c(y)\delta \bar\psi_c(x)}i\tau_2\psi_c(x)
+{1\over g'}\partial_{\mu}
{\delta^3 \Gamma [\phi]\over \delta \bar\psi_c(z)
\delta \psi_c(y)\delta B^{\mu}_c(x)}
-2\epsilon_{2ab}{\delta^3 \Gamma [\phi]\over \delta \bar\psi_c(z)
\delta \psi_c(y)\delta \sigma^a_c(x)}\sigma^b_c(x)=0.
\label{c2}
\end{eqnarray}
%\narrowtext
In the absences of the external sources
\begin{equation}
\psi_c(x)=\bar\psi_c(x)=0,
 \end{equation}
eq. (\ref{c2})  reduces to 
\begin{eqnarray}
&&\delta (x-z)i\tau_2{\delta^2 \Gamma [\phi]\over \delta \psi_c(y)
\delta \bar\psi_c(x)}
+\delta (x-y){\delta^2 \Gamma [\phi]\over \delta\bar \psi_c(z)
\delta \psi_c(x)}i\tau_2+\nonumber\\
&&+2\epsilon_{2ab}{\delta^3 \Gamma [\phi]\over \delta \bar\psi_c(z)
\delta \psi_c(y)\delta \sigma^a_c(x)}\sigma^b_c(x)-{1\over g'}\partial_{\mu}
{\delta^3 \Gamma [\phi]\over \delta \bar\psi_c(z)
\delta \psi_c(y)\delta B^{\mu}_c(x)}
=0.
\label{c3}
\end{eqnarray}
Making Fourier transformation, we can express (\ref{c3}) as
\begin{eqnarray}
\Gamma^{(2)}_{ \psi,\bar\psi}(p) i\tau_2-i\tau_2\Gamma^{(2)}_{ \psi,\bar\psi}(p+k)
&=&2\epsilon_{2ab}\Gamma^{(3)}_{\psi,\bar\psi;\sigma_a}(p+k,-p;-k)\sigma^b_c
-{i\over g'}k_{\mu}\Gamma^{(3)}_{\psi,\bar\psi; B_{\mu}}(p+k,-p;-k).
\label{c4}
\end{eqnarray}
Since we have extracted the composite field $\sigma^a(x)$ as an
independent degree of freedom, 
there is no pole corresponding to this degree of freedom in the proper 
vertices. If $k_{\mu}\to 0$,
(\ref{c4}) becomes
\begin{eqnarray}
-i\bigl[\tau_2, \Gamma^{(2)}_{ \psi,\bar\psi}(p)\bigr]&=&2
\epsilon_{2ab}\Gamma^{(3)}_{\psi,\bar\psi; \sigma_a}(p,-p;0)
\sigma^b_c.
\label{c5}
\end{eqnarray}

In the limit $ p\to 0 $, (\ref{c5}) is
\begin{eqnarray}
-i\bigl[\tau_2, Z^{-1}_{\psi}m_f\bigr]&=&
2\epsilon_{2ab}\Gamma^{(3)}_{\psi,\bar\psi; \sigma_a}(0,0;0)
\sigma^b_c.
\end{eqnarray}
Since the only nonvanishing condensate is $\langle\bar\psi\tau_3\psi\rangle$, 
the fermion mass can be expressed as 
\begin{eqnarray}
m_f&=&m^0_f+\tau_3\delta m_f.
\end{eqnarray}

 Thus from (15) and (16) we get 
\begin{eqnarray}
\delta m_f&=&-\tau_1 Z_{\psi}
\Gamma^{(3)}_{\psi,\bar\psi; \sigma_1}(0,0;0)
\sigma^3_c.
\end{eqnarray}

\begin{figure}
\vspace{-1.4cm}
\centerline{\epsfig{figure=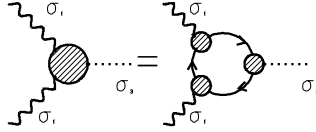,height=6.0cm}}
\vspace{-1.8cm}
\caption{The vertex $\Gamma^{(3)}_{\sigma_1,\sigma_1,\sigma_3}(0,0;0)$. 
\label{fig1}}
\end{figure}

In order to get the mass spectrum of the bound states, we take derivatives 
of (\ref{c1}) with respect to $\sigma_s(y), \sigma_t(z)$
%\widetext
\begin{eqnarray}
&&{1\over g'}\partial_{\mu}
{\delta^2 \Gamma [\phi]\over \delta \sigma_s(y)
\delta B^{\mu}(x)}
+{\delta^2\Gamma [\phi]\over \delta \sigma_s(y)
\delta \psi(x)}i\tau_2\psi(x)
+\bar\psi(x)i\tau_2{\delta^2 \Gamma [\phi]\over\delta \sigma_s(y)
\delta \bar\psi_c(x)}+\nonumber\\
&&-2\epsilon_{2ab}\biggl[\delta_{a,s} \delta (x-y)
{\delta\Gamma [\phi]\over\delta \sigma_b(x)}
+\sigma_a(x){\delta^2\Gamma [\phi]\over\delta \sigma_s(y)
\delta \sigma_b(x)}\biggr]=0,\\
&&{1\over g'}\partial_{\mu}
{\delta^3 \Gamma [\phi]\over \delta\sigma_t(z)\delta \sigma_s(y)
\delta B^{\mu}(x)}
+{\delta^3\Gamma [\phi]\over \delta\sigma_t(z)\delta \sigma_s(y)
\delta \psi(x)}i\tau_2\psi(x)+\nonumber\\
&&+\bar\psi(x)i\tau_2{\delta^2 \Gamma [\phi]\over\delta\sigma_t(z)
\delta \sigma_s(y)\delta \bar\psi_c(x)}
-2\epsilon_{2ab}\biggl[\delta_{a,s} \delta (x-y)
{\delta^2\Gamma [\phi]\over\delta\sigma_t(z)\delta \sigma_b(x)}+\nonumber\\
&&+\delta_{a,t} \delta (x-z)
{\delta^2\Gamma [\phi]\over\delta\sigma_s(y)\delta \sigma_b(x)}
+\sigma_a(x){\delta^3\Gamma [\phi]\over\delta \sigma_t(z)\delta \sigma_s(y)
\delta \sigma_b(x)}\biggr]=0.
\end{eqnarray}
%\narrowtext
From the above equations, we see that the mass spectrum of the bound states 
is 
\begin{mathletters}
\begin{eqnarray}
&m^2_{\sigma_1}&=0,\\
&m^2_{\sigma_3}&=-Z^{-1}_{{\sigma}_1} 
\Gamma^{(3)}_{\sigma_1,\sigma_1;\sigma_3}(0,0;0)\sigma_3,
\label{20b}
\end{eqnarray}
\end{mathletters}
where $\sigma_1$ corresponds to the massless Goldstone boson, and the Feynman 
diagram of $\Gamma^{(3)}_{\sigma_1,\sigma_1;\sigma_3}(0,0;0)$
is shown in Fig.\ (\ref{fig1}). In (\ref{20b}), we have used the relation
\begin{eqnarray}
&Z_{{\sigma}_1}&=Z_{{\sigma}_3},
\end{eqnarray}
which is determined by the O(2) symmetry. As ${\sigma}_1$ is the massless
Goldstone boson, we can get the wavefunction renormalization constant
$Z_{{\sigma}_1}$ from the self-energy
\begin{eqnarray}
&Z_{{\sigma}_1}&={d \over d p^2}\Gamma^{(2)}_{{\sigma}_3}(p)\biggl|_{p^2=0}.
\end{eqnarray}

Similar to the above discussion, taking derivative of (\ref{c1})  with 
respect to $B^{\nu}_c(y)$, we have
\begin{eqnarray}
&&{\delta^2 \Gamma [\phi]\over \delta B^{\nu}_c(y)\delta \psi_c(x)}i\tau_2
\psi_c(x)
+\bar\psi_c(x)i\tau_2{\delta^2 \Gamma [\phi]\over \delta B^{\nu}_c(y)
\delta \bar\psi_c(x)} +\nonumber\\
&&+{1\over g'}\partial^{\mu}{\delta^2 \Gamma [\phi]\over 
\delta B^{\nu}_c(y)\delta B^{\mu}_c(x)}
-2\epsilon_{2ab}\sigma^a_c(x){\delta^2 \Gamma [\phi]\over \delta B^{\nu}_c(y)
\delta \sigma^b_c(x)}
=0.\nonumber\\
\end{eqnarray}
Making Fourier transformation and turning off the external sources, we have 
\begin{equation}
{i\over g'}p_{\mu} \Gamma^{(2)}_{B_{\nu},B_{\mu}}(p)=
-2\sigma^3_c\Gamma^{(2)}_{B_{\nu},\sigma_1}(p).
\label{c6}
\end{equation}
\noindent 
Note that if there is no fermion pair condensate, 
$\langle\bar\psi\tau_3\psi\rangle = 0$ and the symmetry remains, i.e.
\begin{equation}
p_{\mu} \Gamma^{(2)}_{B_{\nu},B_{\mu}}(p)=0.
\end{equation}
\noindent 
This shows that the gauge field $B_{\mu}$ is transverse. If there is
nonvanishing fermion pair condensate, $\langle\bar\psi\tau_3\psi\rangle \ne 0$,
then 
\begin{equation}
p_{\mu} \Gamma^{(2)}_{B_{\nu},B_{\mu}}(p)\ne 0,
\end{equation}
which shows that the $O(2)$ gauge symmetry is dynamically broken.
 
Multiplying $p^{-1}_{\nu}=p_{\nu}/p^2$ to (\ref{c6}) and 
taking $p_{\mu}\to 0$, we have
\begin{equation}
{p_{\mu}p_{\nu}\over p^2} \Gamma^{(2)}_{B_{\nu},B_{\mu}}(p)=
i2{p_{\nu}\over p^2}\Gamma^{(2)}_{B_{\nu},\sigma_1}(p)\sigma^3_c.
\end{equation}
Using the relation
\begin{equation}
\lim_{p\to 0}  \Gamma^{(2)}_{B_{\nu},B_{\mu}}(p)=-Z_B\delta_{\mu\nu}m^2_B,
\label{c12}
\end{equation}
\noindent 
we get the mass of gauge boson $B_{\mu}$
\begin{equation}
m^2_B=-\lim_{p\to 0}Z^{-1}_B g'
i2{p_{\nu}\over p^2}\Gamma^{(2)}_{B_{\nu},\sigma_1}(p)\sigma^3_c.
\label{c9}
\end{equation}
Eq. (\ref{c9}) looks different from the standard formula. In the next
subsection, we shall see that once the matching relation between the
two descriptions is concerned, $m^2_B$ reduces to the standard formula.

\subsection{ Matching Condition}

In the above subsections, we have extracted the composite field $\sigma_a$ 
and treat it as an independent degree of freedom in the effective action.
We regard this as the {\it composite field description}. On the other hand, 
we can also describe the same system by including only the classical
fields of the elementary fields in the effective action, and the bound state
property resides implicitly in the effective action. We regard this as
the {\it elementary field description}.
The two descriptions should be equivalent in describing the same system.
The requirement of equivalence between the two descriptions leads to matching 
conditions for various 1PI vertices which relate the corresponding
quantities in the two descriptions. These conditions are crucially
important in the present approach.

In order to derive the matching conditions, 
we consider now the elementary field description in which the effect of
the external source $K_a$ is implicitly contained in the effective action.
The generating functional in this description is 
\begin{eqnarray}
\bar\Gamma[\bar\phi]&=&W[J]-\int d^4x[\bar\psi_c(x)\eta(x)
+\bar\eta(x)\psi_c(x)+I_{\mu}(x)A^{\mu}_c(x)+J_{\mu}B^{\mu}_c(x)],
\end{eqnarray}
where $\bar\phi\equiv (\bar\psi_c,\psi_c,A^{\mu}_c,B^{\mu}_c,K_a)$. In the 
limit of $K_a=0$, the effective action $\bar\Gamma[\bar\phi]$ 
is just the usual effective action of the fundamental theory.  

Comparing (30) with (\ref{eq14}), we have 
\begin{eqnarray}
\Gamma [\phi]-\bar\Gamma[\bar\phi]&=&\int d^4x\biggl[
\biggl({1 \over a}\sigma_a(x)+\bar\psi_c(x)\tau_a\psi_c(x)\biggr)K_a(x)\biggr].
\end{eqnarray}
\noindent
This is the basic relation between the effective actions in the two
different descriptions. Taking derivatives with respect to the
classical fields, we can obtain the matching conditions for various
1PI vertices.

We first examine the two-point 1PI vertex in the two descriptions. For a given 
$K_a$, taking derivatives of (32) with respect to $\bar\psi_c(x)$ 
and $\psi_c(y)$, we get 
\begin{eqnarray}
{\delta^2(\Gamma[\phi] -\bar\Gamma[\bar\phi])\over \delta\psi_c(y)
\delta\bar\psi_c(x)}\bigg|_{K_a}
=&&-\int d^4x'\biggl[\biggr(
{\delta^2\sigma_a(x')\over \delta\psi_c(y)\delta\bar\psi_c(x)}{1\over a}
\biggr)+\tau_a\delta(x'-y)\delta(x'-x)\biggr] K_a(x).
\end{eqnarray}
In $\Gamma[\phi]$, $\bar{\psi}_c, \psi_c, A^{\mu}_c,$ and $\sigma_a$
are taken to be independent variables. When fixing $K_a$, $\sigma_a$ is
a function of $\bar{\psi}_c,\psi_c$ and $A^{\mu}_c$. Then
%\widetext
\begin{eqnarray}
&&{\delta^2\Gamma[\phi]\over \delta\psi_c(y)
\delta\bar\psi_c(x)}\bigg|_{K_a}-{\delta^2\Gamma[\phi]\over \delta\psi_c(y)
\delta\bar\psi_c(x)}\nonumber\\
&=&\int d^4x_1\biggl[{\delta\sigma_a(x_1)\over\delta\psi_c(y)}
{\delta^2\Gamma[\phi]\over \delta\sigma_a(x_1)\delta\bar\psi_c(x)}
+{\delta^2\Gamma[\phi]\over \delta\sigma_a(x_1)\delta\psi_c(y)}
{\delta\sigma_a(x_1)\over\delta\bar\psi_c(x)}
+{\delta\Gamma[\phi]\over \delta\sigma_a(x_1)}
{\delta^2\sigma_a(x_1)\over\delta\psi_c(y)\delta\bar\psi_c(x)}\biggr]
\nonumber\\
&&+\int d^4x_1d^4x_2{\delta\sigma_a(x_1)\over\delta\psi_c(y)}
{\delta^2\Gamma[\phi]\over \delta\sigma_a(x_1)\delta\sigma_b(x_2)}
{\delta\sigma_b(x_2)\over\delta\bar\psi_c(x)}.
\label{eq39}
\end{eqnarray}
%\narrowtext
Note that in the absence of the external sources, the derivative of 
$\Gamma[\phi]$ with respect to one fermion field $\psi_c(x)$ or 
$\bar\psi_c(x)$ vanishes. Thus when turning off the external sources we have
\begin{eqnarray}
{\delta^2\Gamma[\phi]\over \delta\psi_c(y)
\delta\bar\psi_c(x)}\bigg|_{K_a}&=&{\delta^2\Gamma[\phi]\over \delta\psi_c(y)
\delta\bar\psi_c(x)},
\end{eqnarray}
and 
\begin{eqnarray}
{\delta^2\bar\Gamma[\bar\phi]\over \delta\psi_c(y)
\delta\bar\psi_c(x)}&=&{\delta^2\Gamma[\phi]\over \delta\psi_c(y)
\delta\bar\psi_c(x)},
\label{b4}
\end{eqnarray}
which indicates that the fermion two-point 1PI vertex in the composite field
description is the same as that in the elementary field description.   
This provides a consistent relation between the two descriptions.

\begin{figure}
\vspace{-1.3cm}
\centerline{{\epsfig{figure=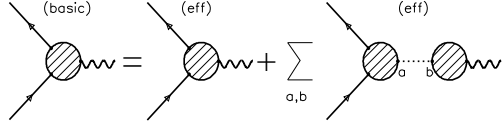,height=5cm}}} 
\vspace{-1.6cm}
\caption{The two-point 1PI vertex in the two descriptions.\label{fig2}}
\end{figure}

Next we look at the three-point 1PI vertex. Similar to the above discussion, 
one can get the Ward-Takahashi identity in the elementary field description
\cite{PTT}. According to (\ref{c4}), we have
\begin{eqnarray}
{k_{\mu}\over 2g'}\bar\Gamma^{(3)}_{\psi,\bar\psi;
B_{\mu}}(p+k,-p;-k)
&=&{k_{\mu}\over 2g'}\Gamma^{(3)}_{\psi,\bar\psi; B_{\mu}}(p+k,-p;-k)
+\epsilon_{2ab}\Gamma^{(3)}_{\psi,\bar\psi;\sigma_a}(p+k,-p;-k)
\sigma_b .
\label{43}
\end{eqnarray}
Differentiating (\ref{eq39}) and (\ref{b4}) with respect 
to $B^{\mu}_c(z)$ and in the limit $I,J\to 0$, we have
\begin{eqnarray}
&&{\delta^3(\Gamma[\phi] -\bar\Gamma[\bar\phi])\over \delta B^{\mu}_c(z)
\delta\psi_c(y)\delta\bar\psi_c(x)}\bigg|_{K_a=0}=0,\\
&&{\delta^3\Gamma[\phi] \over \delta B^{\mu}_c(z)
\delta\psi_c(y)\delta\bar\psi_c(x)}\bigg|_{K_a}-
{\delta^3\Gamma[\phi] \over \delta B^{\mu}_c(z)
\delta\psi_c(y)\delta\bar\psi_c(x)}\nonumber\\
&=&\int d^4x_1d^4x_2
{\delta^2\Gamma[\phi]\over \delta
B^{\mu}_c(z)\delta\sigma_a(x_1)}
D_{ab}(x_1,x_2){\delta^3\Gamma[\phi]\over\delta\psi_c(y)
\delta\bar\psi_c(x)\delta\sigma_b(x_2)}.
\end{eqnarray}
Therefore
\begin{eqnarray}
i\bar\Gamma^{(3)}_{\psi,\bar\psi; B_{\mu}}(p+k,-p;-k)
&=&i\Gamma^{(3)}_{\psi,\bar\psi; B_{\mu}}(p+k,-p;-k)
+i\Gamma^{(3)}_{\psi,\bar\psi; \sigma_a}(p+k,-p;-k)
iD_{ab}(k)i\Gamma^{(2)}_{B_{\mu};\sigma_b}(k).
\label{47}
\end{eqnarray}
Note that in the elementary field description, when $\bar\Gamma^{(2)}_{\psi,
\bar\psi}(p)$ has a symmetry breaking solution, $\bar\Gamma^{(3)}_{\psi,
\bar\psi; B_{\mu}}(p+k,-p;-k) $ has a pole as 
$k_{\mu}\to 0$. Since there is no pole of elementary fields in the 
1PI vertex, the pole is attributed to a bound 
state\cite{CN}\cite{JJ}. There are two kinds of contribution in the vertex 
$\bar\Gamma^{(3)}_{\psi,\bar\psi; B_{\mu}}(p+k,-p;-k) $: the first comes
from the exchange of the bound state (corresponding to $\sigma_a$); the other 
is the regular term coming from the quantum fluctuations of the
elementary fields which is the same as that in the composite field description
as is shown in Fig.(\ref{fig2}). From the right-hand-side of (\ref{47}), we 
see that the pole comes from the massless Goldstone boson $\sigma_1$. In 
the elementary field description, this pole term implicitly resides in the 
1PI vertex $-2\Gamma^{(3)}_{\psi,\bar\psi; \sigma_1}(p+k,-p;-k)\sigma_3$ as 
shown in (\ref{43}). 

\noindent
Since the two-point vertex $\Gamma^{(2)}_{B_{\mu}; \sigma_1}(k)$
reflects the mixing between $B_{\mu}$ and $\partial_{\mu}\sigma_1$,
it must be of the following Lorentz structure 
\begin{eqnarray}
&&\Gamma^{(2)}_{B_{\mu}; \sigma_1}(k)
\equiv k_{\mu}\Gamma^{(2)}_{B; \sigma_1}(k).
\label{c10}
\end{eqnarray}
\noindent 
Then from the equivalence between the pole terms in the two descriptions 
mentioned above, we obtain
\begin{eqnarray}
2\sigma_3&=& -{1\over g'}\Gamma^{(2)}_{B; \sigma_1}(0).
\end{eqnarray}
Eq. (41) gives the relation between the composite field $\sigma_3$ and
the basic quantity $\Gamma^{(2)}_{B;\sigma_1}$ which is of crucial
importance in the present approach. From (41) we see that if 
$\Gamma^{(2)}_{B;\sigma_1}(0)\neq 0$, 
$\sigma_3$ is nonvanishing and the gauge symmetry is dynamical broken; 
otherwise $\sigma_3=0$ and the gauge symmetry remains unbroken. Note
that this provides a method to calculate the fermion condensate 
parameter $\sigma_3$.

\begin{figure}
\vspace{-1.3cm}
\centerline{\epsfig{figure=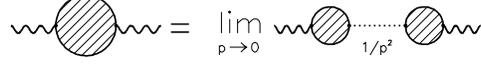,height=5cm}}
\vspace{-1.5cm}
\caption{Dynamical generation of the gauge boson mass.\label{fig3}}
\end{figure}

Inserting the expression of $\sigma_3$ into eq.(\ref{c9}), we get
\begin{eqnarray}
m^2_B=&&\lim_{p\to 0}Z^{-1}_B \Gamma^{(2)}_{B_{\nu},\sigma_1}(p)
{1\over p^2}\Gamma^{(2)}_{B_{\nu}; \sigma_1}(p),
\end{eqnarray}
in which we have used (\ref{c10}), and here $\Gamma^{(2)}_{B; \sigma_1}(p)$ 
is the regular term.  
The gauge boson mass generation can 
be expressed as Fig.(\ref{fig3}).
\noindent This is the dynamical Higgs mechanism\cite{H}.

On the other hand, we can take the composite field description 
to express the gauge field mass as 
\begin{eqnarray}
m_B^2&=&{1\over 4}g'^2F^2_{\pi} ,
\end{eqnarray}
 where 
\begin{equation}
F_{\pi}=4Z^{-1/2}_B\sigma_3.
\end{equation}
This is the familiar gauge boson mass formula in the theory where the 
generated mass results from a non-zero vacuum expectation value of a 
scalar field. 

%\widetext
\subsection{Platform Approximation}

In order to compare our approach with that of Cornwall and Norton, we follow 
them and make the platform approximation\cite{CN}
\begin{mathletters}
\begin{eqnarray}
&\Sigma (p)=&\tau_3\delta m\biggl({p^2\over m^2}\biggr)^{-\epsilon},\\
&Z_{\psi}=&1,\\
&\epsilon =&{3\over (4\pi)^2}(g^2-g'^2),
\end{eqnarray}
\end{mathletters}
\noindent 
where $\delta m=(m_1-m_2)/2$. From (\ref{c4}), we know that
\begin{mathletters}
\begin{eqnarray}
\Gamma^{(3)}_{\psi,\bar\psi; \sigma_1}(p+q,-p;-q)&=&{\tau_1\over 2\sigma_3}
\delta m \biggl[\biggl({p^2\over m^2}\biggr)^{-\epsilon}
+\biggl({(p+q)^2\over m^2}\biggr)^{-\epsilon}\biggr],\\
\Gamma^{(3)}_{\psi,\bar\psi; \sigma_3}(p+q,-p;-q)&=&{\tau_3\over 2\sigma_3}
\delta m \biggl[\biggl({p^2\over m^2}\biggr)^{-\epsilon}
+\biggl({(p+q)^2\over m^2}\biggr)^{-\epsilon}\biggr]
\end{eqnarray}
\end{mathletters}
\begin{figure}
\vspace{-1.3cm}
\centerline{
\epsfig{figure=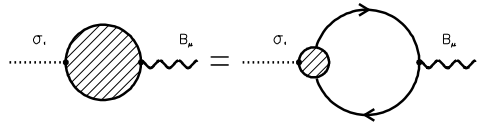,height=5cm}}
\vspace{-1.5cm}
\caption{The gauge boson and composite scalar coupling vertex.\label{fig4}}
\end{figure}

As are shown in Figs. \ref{fig1} and \ref{fig4},
the 1PI vertices $\Gamma^{(2)}_{B_{\mu},\sigma_1}(-q),
\Gamma^{(3)}_{\sigma_1,\sigma_1;\sigma3}(0,0;0)$ and 
$\Gamma^{(2)}_{{\sigma}_1}(q)$ are 
\begin{mathletters}
\begin{eqnarray}
\lim_{q\to 0}i\Gamma^{(2)}_{B_{\mu},\sigma_1}(-q)
&=&\lim_{q\to 0}-{\rm tr}\int {d^4p\over (2\pi)^4}ig'\tau_2\gamma_{\mu}
{1\over\gamma .(p+q)-i\biggl(m+\tau_3\delta m \bigl({(p+q)^2\over m^2}\bigr)^{-\epsilon}
\biggr)}\nonumber\\
&&\quad \times i\Gamma^{(3)}_{\psi,\bar\psi; \sigma_1}(p+q,-p;-q)
{1\over\gamma .p-i\biggl(m+\tau_3\delta m \bigl({p^2\over m^2}\bigr)^{-\epsilon}
\biggr)}\nonumber\\
&=&\lim_{q\to 0}iq_{\mu}g'{1-\epsilon\over \sin 2\pi\epsilon}
{(\delta m)^2\over \sigma_3}{1\over 2\pi}\\
\Gamma^{(3)}_{\sigma_1,\sigma_1;\sigma_3}(0,0;0)&=&
{1-\epsilon\over 2\pi \sin 2\pi\epsilon}
{\delta^4 m\over {\sigma_3}^3}\\
i\Gamma^{(2)}_{\sigma_1}(q)
&=&-{\rm tr}\int {d^4p\over (2\pi)^4}
i\Gamma^{(3)}_{\psi,\bar\psi;\sigma_1}(p,-p-q;q)
{1\over\gamma .(p+q)-i\biggl(m+\tau_3\delta m \bigl({(p+q)^2\over m^2}\bigr)^{-\epsilon}
\biggr)}\nonumber\\
&&\quad \times i\Gamma^{(3)}_{\psi,\bar\psi; \sigma_1}(p+q,-p;-q)
{1\over\gamma .p-i\biggl(m+\tau_3\delta m \bigl({p^2\over m^2}\bigr)^{-\epsilon}
\biggr)}
\end{eqnarray}
\end{mathletters}
%\narrowtext

In the platform approximation, $\epsilon\ll 1, Z_B=1$, so the 
mass spectrum and the Goldstone boson decay constant $F_\pi$ [cf. eq. (44)] 
are 
\begin{mathletters}
\begin{eqnarray}
&m^2_B=&{2\over 3}\biggl({g'^2\over g^2-g'^2}\biggr)(m_1-m_2)^2,\\
&m^2_{\sigma_1}=&0,\\
&m^2_{\sigma_3}=&{(m_1-m_2)^2 \over 4},\\
&\sigma^2_{3}=&{(m_1-m_2)^2\over 32\pi^2\epsilon},\\
&Z_{{\sigma}_1}=&{1\over 32\pi^2\epsilon}
\biggl({{m_1-m_2}\over\sigma_3}\biggr)^2,\\
&F_\pi=&4\sigma_3={{m_1-m_2}\over \sqrt{2\epsilon}\pi}.
\end{eqnarray}
\end{mathletters}
We see that the obtained gauge boson mass $m^2_B$ is exactly the same as 
that in the paper by Cornwall and Norton \cite{CN}, the massless  scalar 
field $\sigma_1$ is the Goldstone boson and the wavefunction renormalization
constant $Z_{\sigma_1}=1$. From Fig.(\ref{fig4}), one can see that the scalar
field $\sigma_1$ is eaten by the gauge boson and becomes the longitudinal 
component of the massive gauge field $B_{\mu}$.  The scalar $\sigma_3$
is a composite Higgs boson with the mass proportional to $m_1-m_2$. 
The Goldstone boson decay constant $F_{\pi}$ is also proportional to $m_1-m_2$ 
or the Higgs boson mass. The results (48c)-(48f) have not been given by
Cornwall and Norton. They are new results from the present approach.

\section{The Jackiw-Johnson Model}
\label{sec:JJ}
\subsection{Ward-Takahashi Identity}

Jackiw and Johnson proposed a model with chiral symmetry which is 
dynamically broken. The model contains a massless fermion field and a 
neutral vector-meson field. 
The Lagrangian density  is 
\begin{eqnarray}
&&{\cal L}=\bar\psi i\gamma\dot\partial\psi
 -{1\over 4}F_{\mu\nu}F_{\mu\nu}
+gJ_{5\mu} A_{\mu},\\
&&J_{5\mu}=i\bar\psi\gamma_{\mu}\gamma_{5}\psi,\\
&&F_{\mu\nu}=\partial^{\mu}A^{\nu}-\partial^{\nu}A^{\mu}.
\end{eqnarray}
As there is an axial-vector coupling, the axial-vector anomaly occurs in 
the theory. The anomaly can be eliminated by introducing additional 
fermions. For the time being, we ignore the effect of anomaly. We 
shall discuss the anomaly and mass generation in the Schwinger model in
the next section.

The Lagrangian is invariant under a $U_A(1)$ chiral transformation
\begin{eqnarray}
&&\delta \psi(x)=i\alpha(x)\gamma_{5}\psi(x),\\
&&\delta A_{\mu}(x)=-{i\over g}\partial_{\mu}\alpha(x),
\end{eqnarray}
We introduce external sources coupling to the composite operators in the
generating functional 
\begin{eqnarray}
Z[J]\equiv &&e^{iW[J]}\nonumber\\
    =&&\int D\bar\psi D\psi DA_{\mu}  
\exp \biggl(i\int d^4x[{\cal L}+J_{\mu}A_{\mu}+\bar\eta\psi
+\bar\psi\eta+\bar{\psi}\psi K+\bar\psi\gamma_5\psi K_5]\biggr).
\end{eqnarray}

We define the classical fields
\begin{mathletters}
\begin{eqnarray}
&&{\delta W[J]\over \delta J_{\mu}(x)}=A^{\mu}_c(x),\\
&&{\delta W[J]\over \delta \bar\eta(x)}=\psi_c(x),\\
&&{\delta W[J]\over \delta \eta(x)}=-\psi_c(x),\\  
&&-{1\over i}{\delta \over \delta \eta(x)}
{\delta \over \delta \bar\eta(x)}W[J]=G(x),\\
&&-{1\over i}{\delta \over \delta \eta(x)}\gamma_5
{\delta \over \delta \bar\eta(x)}W[J]=G_5(x).
\end{eqnarray}
\end{mathletters}
From (54), we have
\begin{eqnarray}
{\delta W[J]\over \delta K_5(x)}=G_5(x)+\bar\psi_c(x)\gamma_5\psi_c(x).
\end{eqnarray}

Under the $U_A(1)$ transformation, we have\footnote{Note that we have
ignored the anomaly.}
%\widetext
\begin{eqnarray}
Z[J]=&&\int D\bar\psi D\psi DA_{\mu} DB_{\mu} 
\biggl[1+i\int d^4x\biggl(-{i\over g}J_{\mu}\partial_{\mu}\alpha(x)
+\bar{\eta}(x)i\gamma_5\alpha(x)\psi(x)+\bar{\psi}(x)i\gamma_5\alpha(x)\eta(x) 
\nonumber\\
&&+2\alpha (x)\bar\psi (x) i\gamma_5\psi(x)K(x)
-2\alpha (x)\bar\psi (x)\psi(x)K_5(x)
\biggr)\biggr]e^{iS_{eff}}.
\end{eqnarray}
Since $\alpha(x)$ is an arbitrary function of $x$, we have
\begin{eqnarray}
&&2K(x)\langle\bar\psi (x) i\gamma_5\psi(x)\rangle
-2\langle\bar\psi (x)\psi(x)\rangle K_5(x)
+\bar\eta(x)i\gamma_5\langle\psi(x)\rangle+
\langle\bar\psi(x)\rangle i\gamma_5\eta(x) 
+{i\over g}\partial_{\mu}J_{\mu}=0,
\label{J1}
\end{eqnarray}

Similar to Sec. \ref{sec:CN}, we take the Legendre transformation
\begin{eqnarray}
\Gamma [\phi]=&&W[J]-\int d^4x[\bar\psi_c(x)\eta(x)+\bar\eta(x)\psi_c(x)+
J_{\mu}(x)A^{\mu}_c(x)\nonumber\\
&&+(G(x)+\bar\psi_c(x)\psi_c(x))K(x)+(G_5(x)+\bar\psi_c(x)i\gamma_5\psi_c(x))K_5(x)]
\end{eqnarray}
and rewrite (\ref{J1}) as
\begin{eqnarray}
&&
{\delta \Gamma [\phi]\over \delta \psi_c(x)}{i\over 2}\gamma_5\psi_c(x)
-\bar\psi_c(x){i\over 2}\gamma_5
{\delta \Gamma [\phi]\over \delta \bar\psi_c(x)}
-{i\over 2g}\partial^{\mu}
{\delta \Gamma [\phi]\over \delta A^{\mu}_c(x)}+
G(x){\delta \Gamma [\phi]\over \delta G_5(x)}
-G_5(x){\delta \Gamma [\phi]\over \delta G(x)}
=0.
\label{J2}
\end{eqnarray}
%\narrowtext
\noindent 
This is the Ward-Takahashi identity in the Jackiw-Johnson model.

Now we introduce the scalar and pseudoscalar composite fields
$\sigma(x)$ and $\pi(x)$, i.e.
\begin{mathletters}
\begin{eqnarray}
 \sigma(x)&=&a G(x),\\
\pi(x)&=&a G_5(x).
\end{eqnarray}
\end{mathletters}
With the composite fields, the Ward-Takahashi identity can be re-expressed as
\begin{eqnarray}
&&
{\delta \Gamma [\phi]\over \delta \psi_c(x)}{i\over 2}\gamma_5\psi_c(x)
-\bar\psi_c(x){i\over 2}\gamma_5
{\delta \Gamma [\phi]\over \delta \bar\psi_c(x)}
-{i\over 2g}\partial^{\mu}{\delta \Gamma [\phi]\over 
\delta A^{\mu}_c(x)}+\sigma_c(x){\delta \Gamma [\phi]\over \delta \pi_c(x)}
-\pi_c(x){\delta \Gamma [\phi]\over \delta \sigma_c(x)}=0.
\label{J3}
\end{eqnarray}
We shall get the mass spectra of fermion, vector meson and scalar
mesons from this Ward-Takahashi identity. 

\subsection{Mass Spectrum}
\widetext
Differentiating (\ref{J3}) with respect 
to $\psi_c(y)$ and $\bar\psi_c(z)$, we have 
\begin{eqnarray}
&&\delta (x-z){i\over 2}\gamma_5{\delta^2 \Gamma [\phi]\over \delta \psi_c(y)
\delta \bar\psi_c(x)}
-\delta (x-y){\delta^2 \Gamma [\phi]\over \delta\bar \psi_c(z)
\delta \psi_c(x)}{i\over 2}\gamma_5
-\bar\psi_c(x){i\over 2}\gamma_5{\delta^3 \Gamma [\phi]\over \delta 
\bar\psi_c(z)\delta \psi_c(y)\delta \bar\psi_c(x)}+\nonumber\\
&&+{\delta^3 \Gamma [\phi]\over \delta \bar\psi_c(z)
\delta \psi_c(y)\delta \bar\psi_c(x)}{i\over 2}\gamma_5\psi_c(x)
-{i\over 2g}\partial_{\mu}
{\delta^3 \Gamma [\phi]\over \delta \bar\psi_c(z)
\delta \psi_c(y)\delta A^{\mu}_c(x)}+\nonumber\\
&&-{\delta^3 \Gamma [\phi]\over \delta \bar\psi_c(z)
\delta \psi_c(y)\delta \sigma_c(x)}\pi_c(x)
+{\delta^3 \Gamma [\phi]\over \delta \bar\psi_c(z)
\delta \psi_c(y)\delta \pi_c(x)}\sigma_c(x)
=0.
\end{eqnarray}
%\narrowtext
\noindent 
We choose the symmetry breaking direction to be
\begin{mathletters}
\begin{eqnarray}
&\langle \bar\psi(x)\psi(x)\rangle &\ne 0,\\
&\langle \bar\psi(x)i \gamma_5\psi(x)\rangle &= 0.
 \end{eqnarray}
\end{mathletters}
Making Fourier transformation we have, for vanishing external sources
\begin{eqnarray}
{i\over 2}\gamma_5\Gamma^{(2)}_{\psi,\bar\psi}(p+k)
+\Gamma^{(2)}_{\psi,\bar\psi}(p) {i\over 2}\gamma_5
&=&{1\over 2g}k_{\mu}\Gamma^{(3)}_{\psi,\bar\psi; A_{\mu}}(p+k,-p;-k)+
\Gamma^{(3)}_{\psi,\bar\psi; \pi}(p+k,-p;-k)
\sigma_c.
\label{J4}
\end{eqnarray}
As $k_{\mu}\to 0$, (\ref{J4}) becomes
\begin{eqnarray}
{i\over 2}\bigl\{\gamma_5, \Gamma^{(2)}_{ \psi,\bar\psi}(p)\bigr\}&=&
\Gamma^{(3)}_{\psi,\bar\psi; \pi}(p,-p;0)
\sigma_c.
\label{J5}
\end{eqnarray}
The fermion mass is now
\begin{eqnarray}
 m_f&=&Z^{-1}_{\psi}\gamma_5
\Gamma^{(3)}_{\psi,\bar\psi; \pi}(0,0;0)
\sigma_c,
\end{eqnarray}
where $Z_{\psi}$ is the wavefunction renormalization constant. 
This is the analogy of the Goldberg-Treiman relation\cite{GT}.

Similar to Sec. \ref{sec:CN}, when we take derivatives of(\ref{J3}) with 
respect to $A_c^{\mu}$, we have 
\begin{eqnarray}
&&\sigma_c\Gamma^{(2)}_{A_{\nu},\pi}(p)-
{i\over 2g}p_{\mu} \Gamma^{(2)}_{A_{\nu},A_{\mu}}(p)=0.
\label{J6}
\end{eqnarray}
From the above equation we see that if chiral gauge symmetry is
unbroken, the gauge field only has transverse components. 
Using the relation (\ref{c12}), we obtain the gauge boson mass
\begin{eqnarray}
&&m^2_A=\lim_{q\to 0} Z_A^{-1}2g{q_{\mu}\over q^2}\Gamma^{(2)}_{A_{\mu},\pi}(q)
\sigma_c,
\label{J7}
\end{eqnarray}
where $Z_A$ is the wavefunction renormalization constant of gauge field.

In  order to determine $\sigma_c$, we can use the matching condition 
\begin{eqnarray}
{k_{\mu}\over 2g}
\bar\Gamma^{(3)}_{\psi,\bar\psi; A_{\mu}}(p+k,-p;-k)
&=&{k_{\mu}\over 2g}
\Gamma^{(3)}_{\psi,\bar\psi; A_{\mu}}(p+k,-p;-k)+
\Gamma^{(3)}_{\psi,\bar\psi; \pi}(p+k,-p;-k)
\sigma_c.
\end{eqnarray}
Since 
\begin{eqnarray}
\bar\Gamma^{(3)}_{\psi,\bar\psi; A_{\mu}}(p+k,-p;-k)
&=&\Gamma^{(3)}_{\psi,\bar\psi; A_{\mu}}(p+k,-p;-k)
+\Gamma^{(3)}_{\psi,\bar\psi; \pi}(p+k,-p;-k)
{-i\over k^2}i\Gamma^{(2)}_{A_{\mu}; \pi}(k),
\end{eqnarray}
we have 
\begin{eqnarray}
&&\sigma_c={1\over 2g}\Gamma^{(2)}_{A; \pi}(0),
\end{eqnarray}
where we have denoted $q_{\mu}\Gamma^{(2)}_{A, \pi}(q)\equiv 
\Gamma^{(2)}_{A_{\mu}, \pi}(q)$. Therefore, we can express the vector meson
mass as
\begin{eqnarray}
m^2_A&=&\lim_{p\to 0}Z^{-1}_A  \Gamma^{(2)}_{A_{\nu},\pi}(p)
{1\over p^2}\Gamma^{(2)}_{A_{\nu}; \pi}(p)
\label{J8},
\end{eqnarray}
which is of the standard form. Introducing the Goldstone boson decay constant 
$F_\pi$, the gauge boson mass can be expressed as 
\begin{eqnarray}
&&m^2_A={1\over 4}g^2F^2_{\pi}\label{J9},\\
&&F_{\pi}=4 Z^{-1/2}_A\sigma_c.
\end{eqnarray}

From the Ward-Takahashi identity, we can also obtain the scalar and 
pseudoscalar meson masses
\begin{mathletters}
\begin{eqnarray}
&&m^2_{\pi}=0,\\
&&m^2_{\sigma}=-Z^{-1}_{\pi}\Gamma^{(3)}_{\sigma,\pi;\pi}(0,0;0)\sigma_c,
\label{J10}
\end{eqnarray}
\end{mathletters}
in which the wavefunction renormalization  constant $Z_{\pi}$ is 

\begin{eqnarray}
&Z_{\pi}&={d \over d p^2}\Gamma^{(2)}_{\pi}(p)\biggl|_{p^2=0}.
\end{eqnarray}
From eqs.(\ref{J4}),(\ref{J5}) and (\ref{J10}), we can express the 1PI vertex
$\Gamma^{(3)}_{\bar\psi,\psi;\pi}$ in terms of 
$\Gamma^{(2)}_{\bar\psi,\psi}$ and obtain all the mass spectrum, the fermion 
condensate and the Goldstone boson decay constant $F_{\pi}$.

In order to compare our results with those in the paper by Jackiw and Johnson, 
we follow them and take the approximation for $\Sigma (p)$\cite{JJ}
\begin{eqnarray}
&&\Sigma (p)= m\biggl({p^2\over m^2}\biggr)^{-\epsilon(g^2)},
\end{eqnarray}
where $\epsilon(g^2)$ is a positive, coupling constant-dependent quantity. 
Then
\begin{mathletters}
\begin{eqnarray}
&&\Gamma^{(2)}_{\psi,\bar\psi} (p)=\gamma .p-i \Sigma(p),\\
&&Z_{\psi}=1,\\
&&Z_A=1.
\end{eqnarray}
\end{mathletters}

Using (\ref{J5}), we get
\widetext
\begin{mathletters}
\begin{eqnarray}
\Gamma^{(3)}_{\psi,\bar\psi;\pi} (p,-p;0)&=&i
\gamma_5{\Sigma(p)\over \sigma_c},\\
i\Gamma^{(2)}_{A_{\mu},\pi}(p)&=&-{\rm tr}\int{d^4q\over (2\pi)^4}\biggl(
{1\over \gamma .q-im\bigl({p^2\over m^2}\bigr)^{-\epsilon}}i\Gamma^{(3)}_{\psi,\bar\psi;\pi}
(p+q,-q;-p)\nonumber\\
& &\quad \times {1\over \gamma .(p+q)-im\bigl({(p+q)^2\over m^2}\bigr)^{-
\epsilon}}ig\gamma_{\mu}\gamma_5
\biggr),\\
i\Gamma^{(3)}_{\sigma,\pi;\pi}(0,0;0)&=&-2{\rm tr}
\int{d^4q\over (2\pi)^4}\biggl(
{1\over \gamma .p-i\Sigma(p)}{\gamma_5\Sigma(p)\over \sigma_c}
{1\over \gamma .p-i\Sigma(p)}\nonumber\\
&&\quad \times{\gamma_5\Sigma(p)\over \sigma_c}
{1\over \gamma .p- i\Sigma(p)}{-i\Sigma(p)\over \sigma_c}\biggr),\\
i\Gamma^{(2)}_{\pi}(q)&=&-{\rm tr}
\int {d^4p\over (2\pi)^4}
\Biggl(i\Gamma^{(3)}_{\psi,\bar\psi;\pi}(p,-p-q;q)
{1\over\gamma .(p+q)-i\biggl(m+\tau_3\delta m \bigl({(p+q)^2\over m^2}
\bigr)^{-\epsilon}\biggr)}\nonumber\\
&&\quad \times i\Gamma^{(3)}_{\psi,\bar\psi; \pi}(p+q,-p;-q)
{1\over\gamma .p-i\biggl(m+\tau_3\delta m \bigl({p^2\over m^2}\bigr)^{-
\epsilon}\biggr)}\Biggr).
\end{eqnarray}
\end{mathletters}
%\narrowtext
Then, we have the results of the mass spectrum and the Goldstone boson decay 
constant
\begin{mathletters}
\begin{eqnarray}
&m^2_A&={g^2\over 4\pi^2}{m^2\over \epsilon(g)},\\
&m^2_{\pi}&=0,\\
&m^2_{\sigma}&=2m^2,\label{J1a}\\
&F_{\pi}&={m\over \pi\epsilon(g)^{1/2}},\\
&Z_{\pi}&={1\over (4\pi)^2\epsilon}\biggl( {m\over \sigma_c}\biggr)^2
\end{eqnarray}
\end{mathletters}
where $m$ is the dynamical fermion mass and $m_{\sigma}$ is the composite Higgs
boson mass. The gauge boson mass is exactly the same as that 
in the paper by Jackiw and Johnson. The composite Higgs boson mass and 
$\pi$ decay constant $F_{\pi}$ are our new results. 
From the above equations, it is easy to see that $Z_{\pi}=1$.
%Note that in the dynamically broken global chiral symmetry theory, such as
%Nambu-Jona-Lasinio and Gross-Neveu models, the mass of the composite 
%Higgs scalar is twice of the fermion mass \cite{Nambu,GN,R}, 
%while in the Jackiw-Johnson model, the composite Higgs 
%scalar mass is $\sqrt{2}m$. Physically, it is reasonable that the composite
%scalar mass is smaller than the sum of individual constitute fermion masses. 
%From (\ref{J10}), we see that the composite scalar mass depends on the
%dynamics in the theory. Through the investigation of composite scalars, 
%we may get information about the dynamics in the theory. 
%The methods, which are only 
%based on symmetry analysis, can't give any prediction of the mass spectrum of 
%composite scalar excitations. 

\section{The Schwinger Model}
\label{sec:S}

In the last section, we have ignored the chiral anomaly, i.e. ignoring 
the variation of the measure in the path integral under the chiral 
transformation.
If a theory is not anomaly free, the integral measure is not invariant under 
the chiral transformation and the Ward-Takahashi identity will be modified.
%Usually, the anomaly will lead some 
%anomaly decays in four dimensions. However, 
In 1+1 dimensions, it has been shown that the chiral anomaly affects the mass 
spectrum \cite{P}. 
For example, in the Schwinger model, the vector boson acquires a mass from
the anomaly. Since the Schwinger model
is exactly solvable, it has been extensively studied in different approaches,
such as functional method, operator method and bosonization techniques.
In this section, we apply  the Ward-Takahashi identity to study
the generation of the vector boson mass.

The Schwinger model is described by the Lagrangian
density
\begin{eqnarray}
{\cal L}&=&\bar\psi i\gamma .D\psi
 -{1\over 4}F_{\mu\nu}F_{\mu\nu}
\end{eqnarray}
where 
\begin{mathletters}
\begin{eqnarray}
&&D_{\mu}=\partial_{\mu}-ie A_{\mu}\\
&&F_{\mu\nu}=\partial_{\mu}A_{\nu}-\partial_{\nu}A_{\mu}
\end{eqnarray}
\end{mathletters}

The generating functional is 
\begin{eqnarray}
Z[J]\equiv &&e^{iW[J]}\nonumber\\
    =&&\int D\bar\psi D\psi DA_{\mu} 
\exp \biggl(i\int d^2x[{\cal  L} +J_{\mu}A_{\mu}+\bar\eta\psi+\bar\psi\eta]\biggr).
\end{eqnarray}

We define the classical fields
\begin{mathletters}
\begin{eqnarray}
&&{\delta W[J]\over \delta J_{\mu}(x)}=A^{\mu}_c(x),\\
&&{\delta W[J]\over \delta \bar\eta(x)}=\psi_c(x),\\
&&{\delta W[J]\over \delta \eta(x)}=-\psi_c(x).
 \end{eqnarray}
\end{mathletters}
Taking the Legendre transformation, we can obtain the 
effective action
\begin{eqnarray}
\Gamma [\phi]&=&W[J]-\int d^2x[\bar\psi_c(x)\eta(x)
+\bar\eta(x)\psi_c(x)
+J_{\mu}(x)A^{\mu}_c(x)].
\end{eqnarray}

Under the chiral gauge transformation
\begin{mathletters}
\begin{eqnarray}
&&\delta \psi(x)=i\gamma_5\beta(x)\psi(x),\\
&&\delta\bar \psi(x)=i\bar \psi(x)\gamma_5\beta(x),\\
&&\delta A_{\mu}(x)={i\over e}\epsilon_{\mu\nu}\partial_{\nu}\beta(x),
\end{eqnarray}
\end{mathletters}
the Lagrangian is invariant and the generating functional is 
\widetext
\begin{eqnarray}
Z[J]&=&\int D\bar\psi' D\psi' DA'_{\mu} 
\exp \biggl(i\int d^2x[{\cal L}+J_{\mu}A'_{\mu}
+\bar\eta\psi'+\bar\psi'\eta]\biggr)\nonumber\\
&=&\int D\bar\psi D\psi DA_{\mu} 
\exp \biggl(i\int d^2x\biggl[{i e\over 2\pi}
\epsilon_{\mu\nu}F_{\mu\nu}\beta(x)+J_{\mu}\delta A_{\mu}
+\bar\eta\delta\psi+\delta\bar\psi\eta\biggr]\biggr) e^{iS_{eff}},
\label{s1}
\end{eqnarray}
where we have considered the variation of the integral measure 
\begin{eqnarray}
 D\bar\psi' D\psi' DA'_{\mu} 
&=& D\bar\psi D\psi DA_{\mu} 
\exp \biggl(-\int d^2x{ e\over 2\pi}\epsilon_{\mu\nu}F_{\mu\nu}\beta(x)
\biggr).
\end{eqnarray}
%\narrowtext

From (\ref{s1}), we can obtain 
\begin{eqnarray}
&&i{e\over \pi}\epsilon_{\mu\nu}\partial_{\mu}A_{c \nu}(x)
-{i \over e}\epsilon_{\mu\nu}\partial_{\nu}J_{\mu}(x) +\bar\eta(x)i\gamma_5\psi_c(x)+\bar\psi_c(x)i\gamma_5
\eta(x)=0.
\label{s2}
\end{eqnarray}
This is the Slavnov-Taylor identity including the effect of the anomaly.
Expressing the external sources in terms of the derivatives of the effective 
action, we get the Ward-Takahashi identity

\begin{eqnarray}
&&i{e\over \pi}\epsilon_{\mu\nu}\partial_{\mu}A_{c \nu}(x)
+{i \over e}\epsilon_{\mu\nu}\partial_{\nu}{\delta \Gamma[\phi]\over
\delta A_{c \mu}(x)}
+{\delta \Gamma[\phi]\over
\delta \psi_c(x)} i\gamma_5\psi_c(x)
-\bar\psi_c(x)i\gamma_5{\delta \Gamma[\phi]\over
\delta\bar\psi_c(x)} =0.
\label{s3}
\end{eqnarray}
Taking derivative with respect to $A_{c \lambda}(y)$, we have 
\begin{eqnarray}
&&{\delta^2 \Gamma[\phi]\over\delta A_{c\lambda}(x)
\delta \psi_c(x)} i\gamma_5\psi_c(x)
+\bar\psi_c(x)i\gamma_5
{\delta^2 \Gamma[\phi]\over\delta A_{c \lambda}(x)
\delta\bar\psi_c(x)}+ \nonumber\\
&&+i{e\over \pi}\epsilon_{\mu\nu}\partial_{\mu}\delta(x-y)\delta_{\nu\lambda}
+{i \over e}\epsilon_{\mu\nu}\partial_{\nu}{\delta^2 \Gamma[\phi]\over
\delta A_{c\lambda}(y)\delta A_{c \mu}(x)} =0.\nonumber\\
\label{s4}
\end{eqnarray}
In the absence of the external sources, (\ref{s4}) becomes
\begin{eqnarray}
&&i{e^2\over \pi}k_{\mu}k_{\mu}+ik_{\mu}k_{\nu}\Gamma^{(2)}_{\mu\nu}(k)=0.
\end{eqnarray}
Thus the vector boson mass is 
\begin{eqnarray}
m^2_A=&&-\lim_{k\to 0}{k^{\mu}k^{\nu}\over k^2}\Gamma^{(2)}_{\mu\nu}(k)
\nonumber\\
=&&{e^2\over \pi}.
\end{eqnarray}
This coincides with the results in other approaches. 
In this approach, we see clearly that, in the Schwinger model, the vector 
meson acquires a mass from the anomaly related to the measure of the
path integral as it should be. 

\section{Conclusions}
\label{sec:D}

In this paper, we have developed a formal approach to dynamical
breaking of Abelian gauge symmetry based on the Ward-Takahashi identity
including composite fields and the matching condition for the
equivalence between the elementary field description and the composite
field description. The approach is easily applied to the Cornwall-Norton 
model,the Jackiw-Johnson model and the Schwinger model. In this
approach, we can obtain not only the dynamically generated gauge boson 
and fermion masses as in other approaches, but also the masses of the 
composite scalars and the Goldstone boson decay constant $F_\pi$ which
have not been given in the corresponding papers.

Of course, the explicit results of the mass spectra and the Goldstone
boson decay constant depends on the evaluation of the 1PI vertices 
$\Gamma^{(3)}_{\psi,\bar\psi;\pi}$ and $\Gamma^{(3)}_{\psi,\bar\psi;\sigma_1}$, 
which requires certain approaches to the dynamics. In this paper, for the sake 
of comparison with the existing results, we simply take the simple 
approximations following the corresponding papers. Developing further improved 
approximations will be of interest in future investigations.

The approach developed in this paper can in principle be generalized to
the study of dynamical symmetry breaking in non-Abelian gauge theories.

\section*{Acknowledgments}
One of us (K.S)  would like to thank Prof. Qing Wang for 
many enlighting discussions and advice, and is grateful to 
Profs. Zhongping Qiu and Yu-Ping Yi for helpful discussions. 
This work is partly supported by the National Natural Science Foundation 
of China, the National Post-doctoral Science Foundation of China, the 
Fundamental Research Foundation of Tsinghua University,
, and a special grant from the State Commission of Education of China.

\end{document}